
\magnification=1200
\tolerance=5000
\pretolerance=500
\vsize=24.5truecm
\hsize=17.0truecm      \hoffset=-0.5truecm
\baselineskip=16pt
\nopagenumbers

\def\square{\hbox{\rlap{$\sqcap$}$\sqcup$}}

\def\alt{\hbox{\raise.5ex\hbox{$<$}
\kern-1.1em\lower.5ex\hbox{$\sim$}}}
\def\agt{\hbox{\raise.5ex\hbox{$>$}
\kern-1.1em\lower.5ex\hbox{$\sim$}}}

\null \vskip0.5truecm
\centerline{{\bf STRING THEORY AND GRAVITY}\footnote{*}
{First award of the Gravity Research Foundation,1994;
to appear in Gen.Rel.Grav. (1994)} }
\medskip\medskip
\centerline{T. Damour}
\centerline{Institut des Hautes Etudes Scientifiques}
\centerline{ 91440 Bures sur Yvette, France }
\medskip \smallskip
\centerline{A.M. Polyakov}
\centerline{Physics Department, Princeton University}
\centerline{Princeton, New Jersey 08544, U.S.A.}
\vskip8.5truecm
{   \baselineskip=12pt
It is pointed out that string-loop effects may generate matter couplings
for the dilaton allowing this scalar partner of the tensorial graviton
to stay massless while contributing to macroscopic gravity in a way naturally
compatible with existing experimental data. Under a certain assumption
of universality of the dilaton coupling functions (possibly realized
through a discrete symmetry such as S-duality), the cosmological
evolution drives the dilaton towards values where it decouples from
matter. At the present cosmological epoch, the coupling to matter of the
dilaton should be very small, but non zero. This provides a new
motivation for improving the experimental tests of Einstein's
Equivalence Principle. \par
}
\vfill \eject

String theory is, at present, the only scheme promising to provide a
combined quantum theory of gravity and of the gauge interactions. It is
striking that, within string theory, the usual Einsteinian tensor
graviton is intimately mixed with a scalar partner: the dilaton. The
matter couplings of the dilaton {\it a priori} generate drastic
deviations from general relativity, notably violations of Einstein's
Equivalence Principle (EP): universality of free fall, constancy of the
constants,\dots This is why it is generally assumed that the dilaton
will acquire a (Planck scale) mass due to some yet unknown dynamical
mechanism.

\medskip
An alternative possibility is the following (see [1]
for a detailed discussion) : string-loop
effects (associated with worldsheets of arbitrary genus in intermediate
string states) may naturally reconcile the existence of a massless
dilaton with existing experimental data if they exhibit a certain kind of
universality.

\medskip
To illustrate this possibility, let us consider a toy model
where the effective action for the string
massless modes (considered directly in four dimensions) takes the form
$$
S = \int d^4 x\, \sqrt{\hat g}\, B (\Phi) \left\{ {1\over \alpha '}
 [\widehat R + 4 \widehat{\square} \Phi - 4 (\widehat\nabla \Phi)^2]
 - {k\over 4} \widehat F^2 - \overline{\widehat \Psi}\hat D \widehat \Psi
  +\dots \right\}\ , \eqno(1)
$$
\noindent
where the function of the dilaton $\Phi$ appearing as a common factor in
front is given by a series of the type
$$ B(\Phi) = e^{-2\Phi} + c_0 + c_1\, e^{2\Phi} + c_2\, e^{4\Phi} + \dots
  \eqno(2) $$
The first term on the right-hand side of Eq.~(2) is the string tree
level contribution (spherical topology for intermediate worldsheets)
which is known to couple in a universal multiplicative manner [2]
[3] [4]. The further terms represent the string-loop effects:
the genus-$n$ string-loop contribution containing a factor $g^{2(n-1)}_s$
where $g_s\equiv \exp (\Phi)$ is the string coupling constant. Apart
from the fact that Eq.~(2) is a series in powers of $g^2_s$, little
is known about the global behaviour of the dilaton coupling function
$B(\Phi)$.
In general, one expects to have several different coupling functions
$B_i(\Phi)$ as coefficients of the terms in the action.
 For the cosmological attractor mechanism discussed here to apply
the following universality condition must be fulfilled:
the various coupling functions of the dilaton $B_i(\Phi)$ must all
admit a local maximum at some common value of $\Phi$.
 In the toy model illustrated in Eq.(1),
this universality is guaranteed by the factorization of a common function
 $B(\Phi)$; string-loop effects, Eq.(2), can then allow this function
to admit a local maximum.  More general ways of realizing the
needed universality are discussed in [1]. In particular, it is suggested
in [1] that a discrete symmetry, e.g. under $\Phi \rightarrow -\Phi$,
may be all what is needed to ensure the minimal required universality.
We note that precisely this symmetry (i.e. $g_s \rightarrow 1/g_s$)
has been conjectured to hold for the dilaton in string theory ( S-duality).
Another possibility is that our attractor mechanism could apply to the other
gauge-neutral massless scalar fields present in string theory ( moduli),
which are known to possess such a discrete symmetry ( target-space duality).
For simplicity, we shall phrase our results in terms of the toy model (1).

It is convenient to transform the action (1) by introducing several
$\Phi$-dependent rescalings. In particular, one replaces the original
``string-frame" metric $\hat g_{\mu\nu}$ by a conformally related
``Einstein-frame" metric $g_{\mu\nu} \equiv C \, B(\Phi) \hat
g_{\mu\nu}$, and the original dilaton field $\Phi$ by a canonical scalar
field $\varphi$. The transformed action reads
$$ S = \int d^4 x \sqrt g \left\{ { 1\over 4q} R - {1\over 2q} (\nabla
\varphi)^2 - \overline\Psi D \Psi - {k\over 4} B(\varphi) F^2 + \dots
\right\}\ , \eqno(3) $$
\noindent
where $q \equiv 4\pi \overline{G} \equiv {1\over 4} C\alpha'$ denotes
a bare gravitational coupling constant and $B(\varphi) \equiv B [\Phi
(\varphi)]$.

The basic clue which allows one to relate string models to the observed
low-energy world is the dilaton dependence of the gauge coupling
constants apparent in Eq.~(3): $g^{-2} = kB (\varphi)$. This dependence
implies that the QCD mass scale $\Lambda_{QCD}$ is given by
$$ \Lambda_{QCD} (\varphi) \sim C^{-1/2} B^{-1/2} (\varphi) \exp [-8
   \pi^2 b^{-1}_3 k_3 B(\varphi)] \widehat \Lambda _s \ , \eqno(4) $$
\noindent where $b_3$ is a (rational) one-loop coefficient associated
with the scale dependence of the SU(3) coupling constant, and where
$\widehat\Lambda_s \simeq 3 \times 10^{17}$GeV [5] is the (string-frame)
string unification scale $\propto \alpha '^{-1/2}$. The Einstein-frame
mass of hadrons is essentially some pure number times
$\Lambda_{QCD}(\varphi)$.  More generally, under the assumption of a
universal $B(\varphi)$, the masses of all the particles will depend on
$\varphi$ only through the function $B(\varphi)$:
$$ m_A (\varphi) = m_A [B(\varphi)]\ . \eqno(5) $$
When studying the cosmological evolution of the graviton-dilaton-matter
system, one finds that the dilaton vacuum expectation value $\varphi$
is dynamically driven toward the values $\varphi_m$ corresponding to
a local maximum of $B(\varphi)$, i.e.
 a local minimum of all the various mass
functions $m_A (\varphi)$. [With some important physical differences,
this cosmological attractor mechanism is similar to the one discussed in
Ref.~[6] which concerned metrically-coupled tensor-scalar theories].
The main parameter determining the efficiency of the cosmological relaxation
of $\varphi$ toward $\varphi_m$ is the curvature $\kappa$ of the
function $\ln B (\varphi)$ near the maximum $\varphi_m$:
$$ \ln B(\varphi) \simeq \ {\rm const.} \ - {1\over 2} \kappa
 (\varphi - \varphi_m)^2 \ . \eqno(6) $$

Because of the steep dependence of $m_A (\varphi)$ upon $B(\varphi)$
[illustrated by Eq.~(4)], each ``mass threshold'' during the
radiation-dominated era [i.e. each time the cosmic temperature $T$
becomes of order of the mass $m_A$ of some particle] attracts
 $\varphi$  towards  $\varphi_m$
by a factor $\sim 1/3$ . In the subsequent
matter-dominated era, $\varphi$ is further attracted toward $\varphi_m$
by a factor proportional to $Z^{-3/4}_0$ where $Z_0 \simeq 1.3 \times
10^4$ is the redshift separating us from the end of the radiation era.
Finally, in the approximation where the phases of the ten or so
successive relaxation oscillations around $\varphi_m$ undergone by
$\varphi$ during the cosmological expansion are randomly distributed,
one can estimate (when $\kappa \agt 0.5$)
that the present value $\varphi_0$ of the dilaton
differs, in a $rms$ sense, from $\varphi_m$ by
$$ (\varphi_0 -\varphi_m)_{rms} \sim 2.75 \times 10^{-9} \times
 \kappa^{-3} \Omega^{-3/4}_{75} \Delta \varphi\ , \eqno(7) $$
where $\Omega_{75} \equiv \rho^{\rm matter}_0 / 1.0568 \times 10^{-29}$%
g~cm$^{-3}$  and where $\Delta\varphi$ denotes the deviation of $\varphi$
from $\varphi_m$ at the beginning of the (classical) radiation era.

The  present scenario  predicts the existence of many small, but non
zero, deviations from general relativity. Indeed, a cosmologically
relaxed dilaton field couples to matter around us with a strength
(relative to usual gravity)
$$ \alpha_A = \left. {\partial \ln m_A (\varphi)\over \partial\varphi}
  \right|_{\varphi_0} \simeq \beta_A (\varphi_0 - \varphi_m) \eqno(8) $$
with $\beta_A \simeq \beta_3 \equiv 40.8 \kappa$ for hadronic matter.
Therefore, all deviations from Einstein's theory contain a small factor
$(\varphi_0 -\varphi_m)^2$ coming from the exchange of a $\varphi$
excitation. More precisely, the post-Newtonian deviations from general
relativity at the present epoch are given by the Eddington parameters
$$ \eqalignno{
1 -\gamma_{\rm Edd} &\simeq 2 (\beta_3)^2 (\varphi_0 -\varphi_m)^2\ , &(9)\cr
\beta_{\rm Edd} -1 &\simeq {1\over 2} (\beta_3)^3 (\varphi_0 -\varphi_m)^2
  \ , & (10) \cr }   $$
while the residual cosmological variation of the coupling constants is
at the level
$$ \eqalignno{
 {\dot\alpha\over \alpha} & \simeq - \kappa \left[ \omega \tan \theta_0
    + {3\over 4} \right] (\varphi_0 - \varphi_m)^2 H_0\ , & (11) \cr
 {\dot G\over G} & \simeq - 2 \beta^2_3 \left[ \omega \tan \theta_0
    + {3\over 4} \right] (\varphi_0 - \varphi_m)^2 H_0\ , & (12) \cr }$$
where $\omega \equiv \left[ {3\over 2} (\left( \beta_3 - {3\over 8}
\right) \right]^{1/2}$, and where $\theta_0$ denotes the phase of the
matter-era relaxation toward $\varphi_m$, while $H_0$ denotes the present
value of Hubble's ``constant''.

The most sensitive way to look for the existence of a weakly coupled
massless dilaton is through tests of the universality of free fall.
The interaction potential between particle $A$ and particle $B$ is
$-G_{AB} m_A m_B /r_{AB}$ where $G_{AB} = \overline G (1+\alpha_A
\alpha_B)$. Therefore two test masses, $A$ and $B$, will fall in the
gravitational field generated by an external mass $m_E$ with
accelerations $a_A$ and $a_B$ differing by
$$ \left( {\Delta a\over a} \right)_{AB}\equiv 2\ {a_A -a_B\over a_A +a_B}
  \simeq (\alpha_A - \alpha_B) \alpha_E\ . \eqno(13) $$
The difference $\alpha_A - \alpha_B$ introduces a small factor proportional
to the ratio $m_{\rm quark}/ m_{\rm nucleon}$ or to the fine structure
constant $\alpha$. Finally, one finds an equivalence-principle
violation of the form
$$ \left( {\Delta a\over a}\right)_{AB} = \kappa^2 (\varphi_0 -\varphi_m)^2
 \left[ C_B \Delta \left({B\over M}\right)
  + C_D \Delta \left({D\over M}\right)
  + C_E \Delta \left({E\over M}\right) \right]_{AB} \ , \eqno(14) $$
where $B \equiv N+Z$ is the baryon number, $D\equiv N-Z$ the neutron
excess, $E\equiv Z (Z-1)/(N+Z)^{1/3}$ a Coulomb energy factor and $M$
the mass of a nucleus. The only coefficient in Eq.~(14) which can be
reliably estimated is the last one: $C_E \simeq 3.14 \times 10^{-2}$.
The largest $\Delta a/a$ will arise in comparing Uranium $(E/M \simeq
5.7)$ with Hydrogen (or some other light element). For such a pair
Eqs.~(7) and (14) yield a violation of the equivalence principle
which is well below the present experimental limits
$$ \left( {\Delta a\over a} \right)^{\max}_{rms} = 1.36 \times 10^{-18}
  \kappa^{-4} \Omega^{-3/2}_{75} (\Delta \varphi)^2\ . \eqno(15) $$

 The results (9)-(12) and (14),(15) provide a new motivation for trying
to improve by several orders of magnitude the experimental tests of
general relativity, notably the tests of the equivalence principle
(universality of free fall, constancy of the constants,\dots).
Let us note that if our mechanism were to apply to a modulus field ,
rather than to the dilaton, the final observational deviations from
General Relativity would be expected to be numerically more important
because of the less steep dependence of the masses as functions of
a modulus.

The scenario summarized here gives an example of a
well-motivated theoretical model containing  no small
parameters and naturally predicting very small deviations from general
relativity at the present epoch. In this model, high-precision tests of
the equivalence principle can be viewed as low-energy windows on
string-scale physics: not only could they discover the dilaton, but, by
measuring the ratios $C_B/C_E$, $C_D/C_E$ in Eq.~(14) they would probe
some of the presently most obscure aspects of particle physics: Higgs
sector and unification of coupling constants.

\bigskip

REFERENCES

\bigskip
{ \baselineskip=12pt
\item{[1]}T. Damour and A.M. Polyakov, Nucl. Phys. B.{\bf 423},532 (1994).
\item{[2]}E.S. Fradkin and A.A. Tseytlin, Phys. Lett. B{\bf 158}, 316 (1985).
\item{[3]}C.G. Callan, D. Friedan, E.J. Martinec and M.J. Perry, Nucl. Phys.
B{\bf 262}, 593 (1985).
\item{[4]}C.G. Callan, I.R. Klebanov and M.J. Perry, Nucl. Phys. B{\bf 278},
78 (1986).
\item{[5]}V.S. Kaplunovsky, Nucl. Phys. B{\bf 307}, 145 (1988).
\item{[6]}T. Damour and K. Nordtvedt, Phys. Rev. Lett. {\bf 70}, 2217 (1993);
 Phys. Rev. D {\bf 48}, 3436 (1993).
\par }
\bye